\documentclass[conference]{IEEEtran}
\usepackage{bm}
\IEEEoverridecommandlockouts
\normalsize

\usepackage{url}


\usepackage[utf8]{inputenc}
\usepackage[T1]{fontenc}
\usepackage{mathptmx} 

\usepackage{soul}
\usepackage{array}
\usepackage{color}
\usepackage{amsfonts}
\usepackage{amssymb}
\usepackage{amsmath}
\usepackage[pdftex]{graphicx}
\usepackage{cite}
\usepackage{balance}
\usepackage{caption}
\usepackage{subcaption}
\usepackage{textcomp}
\usepackage{gensymb}
\usepackage{float}
\usepackage{comment}
\usepackage{subcaption}
\usepackage{tabularx}
\usepackage{multirow}
\usepackage{amsthm}
\usepackage{algorithmic}
\usepackage{algorithm}
\usepackage{booktabs}

\usepackage[utf8]{inputenc}
\usepackage{mathptmx}
\IEEEoverridecommandlockouts
\begin{document}
\setlength{\columnsep}{0.26in}


\title{Digital-Twin Empowered Site-Specific Radio Resource Management in 5G Aerial Corridor}
\author{Pulok Tarafder$^{\dagger}$, Imtiaz Ahmed$^{\dagger}$, Danda B. Rawat$^{\dagger}$, Md. Zoheb Hassan$^{\dagger\dagger}$, and Kamrul Hasan$^{\dagger\dagger\dagger}$\\
$^{\dagger}$Howard University, Washington, DC, USA, $^{\dagger\dagger}$University of Laval, Quebec City, QC, Canada, \\$^{\dagger\dagger\dagger}$Tennessee State University, Nashville, TN, USA\\
Emails: pulok.tarafder@bison.howard.edu, imtiaz.ahmed@howard.edu, danda.rawat@howard.edu, \\ md-zoheb.hassan@gel.ulaval.ca,  mhasan1@tnstate.edu.}

\maketitle

\begin{abstract}
Base station (BS) association and beam selection in multi-cell drone corridor networks present unique challenges due to the high altitude, mobility and three-dimensional movement of drones. These factors lead to frequent handovers and complex beam alignment issues, especially in environments with dense BS deployments and varying signal conditions. To address these challenges, this paper proposes a channel-twin (CT) enabled resource-allocation framework for drone-corridor communications, where the CT constitutes the radio-channel component of a broader digital-twin (DT) environment. The CT supplies high-fidelity channel-state information (CSI), which drives a two-stage optimization procedure. In Stage 1, array-level beamforming weights at each BS are selected to maximize antenna gain. In Stage 2, the framework jointly optimizes drone–BS–beam associations at discrete corridor way-points to maximize end-to-end throughput. Simulation results confirm that the CT-driven strategy delivers significant throughput gains over baseline methods across diverse operating scenarios, validating the effectiveness of integrating precise digital-twin channel models with cross-layer resource optimization.
\end{abstract}

\vspace{-0.2cm}
\section{Introduction}

Digital-twin (DT) technology has rapidly become a cornerstone of data-driven wireless‐network research, fuelled by recent breakthroughs in artificial intelligence (AI) and machine learning (ML) \cite{lin20236g}. By creating high-fidelity, physics-aware replicas of the propagation environment, DT platforms empower researchers to emulate end-to-end system behaviour, obtain near-real-world performance metrics, and iteratively refine next-generation (NextG) communication algorithms within a controllable virtual space. These virtual networks can remain tightly synchronized with live measurements of the physical networks, furnishing site-specific channel realizations and network parameters while still enabling exploratory ``what-if'' analyses \cite{haider2025digital,elloumi2025spectrum, elloumi2025digital}. Essential capabilities such as city-scale three-dimensional (3D) mapping, multi-modal sensing fusion, and low-latency inference for real-time reconfiguration position, DTs serve as a key enabler for the NextG of communications \cite{alkhateeb2023real}. Leveraging such large-scale, physically grounded digital replicas allows the network operators to streamline radio-access-network (RAN) planning: candidate base-station (BS) locations, antenna topologies, and resource-management policies. Such parameters can be stress-tested virtually, thereby reducing the cost, time, and risk inherent in traditional field-trial-driven deployment cycles.


Unmanned aerial vehicles (UAVs), hereafter referred to collectively as drones, have attracted sustained scholarly attention due to their agility, autonomy, and suitability for a broad range of civilian and commercial missions \cite{beard2012small}. Recent advances in lightweight airframes, propulsion systems, and on-board processing have extended this versatility to wireless networking, enabling platforms that range from quad-rotors to balloons and solar-powered airships to serve as aerial communication nodes \cite{serrano2021balloons}. Moreover, UAVs are rapidly emerging as versatile platforms for diverse applications ranging from search-and-rescue and surveillance to acting as aerial access points for cellular connectivity and enabling last-mile delivery in e-commerce \cite{cherif20213d}. Recognized as a cost-effective and environmentally sustainable solution for commercial package transport, their growing adoption underscores the critical need to investigate structured aerial pathways which is commonly referred to as drone corridors to ensure safe, scalable, and efficient UAV operations.

Within this emerging paradigm, DT environments are essential for studying the unique propagation, mobility, and energy constraints of drone corridor and 3D airspace. By providing physics-consistent channel realizations and accurate beam-management feedback, DTs enable rigorous evaluation of throughput-maximization and link-adaptation strategies before live flight tests. In \cite{karimi2024optimizing}, an optimization framework defines a cellular-connectivity-aware ''UAV corridor'', effectively creating an aerial highway with guaranteed coverage. In \cite{lee2024digital}, a deep-Q-network agent is integrated within a UAV DT to dynamically reposition aerial relays for capacity enhancement. In \cite{chowdhury20203}, an energy-efficient 3D trajectory is derived that leverages 3GPP-compliant beam patterns, while \cite{hazarika2023radit} employs a DT-driven deep-reinforcement-learning scheme to jointly optimize flight paths, energy use, and task off-loading. Collectively, these studies highlight the value of DT-enabled methods for accelerating the design and deployment of robust, high-capacity UAV communication networks.


Although interest in DT-enabled aerial networking is accelerating, the literature still lacks a unified framework that jointly optimizes UAV–BS association and beam selection. Bridging this gap demands an accurate digital replica of the wireless channel along the drone corridor, a component we term the channel twin (CT). A precise CT is the cornerstone for any resource-allocation strategy, as it supplies the physics-consistent link states needed for reliable algorithm design and evaluation. In this work, we propose a CT-driven UAV-assisted downlink architecture tailored to NextG networks. This work makes the following contributions. First,  high-fidelity ray-tracing empowered CT is developed where multiple UAVs are served by terrestrial BSs equipped with directional antenna arrays that conform to ECC Report 281 and 3GPP TR 37.840 specifications \cite{ECCReport281,3gpp-tr-37840}. Second, by leveraging these standard-compliant patterns, we devise a two-stage optimization framework that 1) adapts beamforming weights to maximize array gain, and 2) assigns each UAV to a BS–beam pair at discrete waypoints along predefined aerial corridors to maximize end-to-end throughput. Finally, extensive simulation results are presented to depict the effectiveness of the proposed approach. To the best of our knowledge, this study delivers the first DT benchmark that simultaneously optimizes physical-layer beam design and network-layer UAV association in a NextG context.


\section{System Model}

\noindent\textbf{System Parameters:} We consider a aerial corridor network comprising $M$ receivers (UAVs) and $L$ transmitters (BSs), where each BS is equipped with $N$ transmit antennas arranged in a uniform planar array (UPA). The directional gain toward a given UAV depends jointly on the array beamforming weights and the individual element radiation pattern. Each BS can create at most N directional beams with its transmit power equally divided among all beams. Let the $l$-th BS employ a beamforming codebook $\mathcal{C}_l=\{w_{1,l}, w_{2,l}, \cdots, w_{N,l}\}$, where $w_{n,l}$ is the beamforming vector for the $n$-th beam. We assume that each BS's transmit power is equally divided among all $N$ beams. Each beam can support only one drone, and each drone can be associated with one beam and one BS.



\noindent\textbf{Construction of CT:} The CT is generated using \emph{NVIDIA Sionna}, a GPU-accelerated, open-source link-level simulator that integrates a physics-based ray-tracer built on Mitsuba-3 and TensorFlow \cite{tensorflow2015-whitepaper,hoydis2023sionna,Mitsuba3}. The workflow comprises three stages:

\textbf{1)} \textit{3D environment modelling}: A detailed model of the specific site (in Section IV, we consider the Howard University campus located in Washington, DC, USA) is imported into Blender from OpenStreetMap. Building geometries are projected into three dimensions, and International Telecommunication Union (ITU) material profiles are assigned to all surfaces to obtain realistic reflection, diffraction, and penetration characteristics.

\textbf{2)} \textit{Infrastructure integration}: Geolocated cellular-BS coordinates obtained from the OpenCelliD database \cite{OpenCelliD} are embedded in the scene, providing real-world transmitter positions and antenna heights.  

\textbf{3)} \textit{Ray-tracing and channel extraction}: The annotated scene is loaded into \textit{Sionna}, which traces multipath components between every UAV waypoint and each BS. The resulting complex baseband coefficients are assembled into the channel tensor $\mathbf{H} \in \mathbb{C}^{M \times L \times N}$, where $h_{i,j,k}$ denotes the gain from BS $j$, antenna $k$ to UAV $i$, with $i \in \{1,\dots,M\}$, $j \in \{1,\dots,L\}$, and $k \in \{1,\dots,N\}$.

\noindent\textbf{Transmitter Antenna Gain Model:} Based on the 3GPP TR 37.840 5G specifications, the overall transmitter antenna gain can be modeled as 
\begin{equation}\label{antenna_gain}
    G_{5 \mathrm{G}}\left(\theta, \phi\right)=A_E\left(\theta, \phi\right)+A_V\left(\theta, \phi\right),
\end{equation}
where $A_E\left(\theta, \phi\right)$ and $A_V\left(\theta, \phi\right)$ are the antenna element gain and array gain in dBi unit, respectively. The directional pattern of antenna elements $A_E(\theta,\phi)$ are represented as follows:
\begin{equation}
A_E(\theta,\phi) = G_{E,\text{max}} - \min\{-[A_{E,V}(\theta) + A_{E,H}(\phi)], A_m\}.
\end{equation}
Here, 
\begin{equation}
A_{E,V}(\theta) = -\min\left\{12\left(\frac{\theta - 90^\circ}{\theta_{3\text{dB}}}\right)^2, SL_{A,V}\right\},
\end{equation}
and
\begin{equation}
A_{E,H}(\phi) = -\min\left\{12\left(\frac{\phi}{\phi_{3\text{dB}}}\right)^2, A_m\right\}.
\end{equation}
Moreover, the array beam forming directional pattern gain $A_V(\theta,\phi)$ can be modeled as
\begin{equation}
A_V\left(\theta, \phi\right)=10 \log _{10}\left(\left|\mathbf{V}^H\left(\theta, \phi\right) \mathbf{W}\left(\phi_{\text {scan }}\right)\right|^2\right),
\end{equation}
where $\mathbf{V}^H\left(\theta, \phi\right)$ and $\mathbf{W}\left(\phi_{\text {scan }, i}\right)$ are $N_H \times N_V$-long steering and beamforming vectors, respectively, and $\phi_{\text {scan }}$ is the scanning angle. The ($m,n$)-th elements of $\mathbf{V}^H\left(\theta, \phi\right)$ and $\mathbf{W}\left(\phi_{\text {scan }}\right)$ are expressed as
\begin{equation}
\mathbf{V}^H_{m,n} = \exp\left[\frac{2\pi j}{\lambda} \left((m{-}1)d_H \sin\theta \sin\phi + (n{-}1)d_V \cos\theta\right)\right],
\end{equation}

and
\begin{multline}
\mathbf{W}^H_{m,n} = \frac{1}{\sqrt{N_H N_V}} \exp\Bigg[-j\frac{2\pi}{\lambda}\Big( (m-1)d_H\sin(\phi_{\text{SCAN}})\\ \cos(\theta_{\text{TILT}}) - (n-1)d_V\sin(\theta_{\text{TILT}}) \Big)\Bigg],
\end{multline}
respectively. Here, $N_H$ and $N_V$ denote the number of horizontal and vertical antenna elements, respectively. Furthermore, $d_H$ and $d_V$ represent the horizontal and vertical antenna element spacing, respectively, and $\lambda$ denotes the wavelength. The horizontal beam steering scan angle is represented by $\phi_{\text{SCAN}}$, and $\theta_{\text{TILT}}$ denotes the downtilt beam steering tilt angle. These parameters collectively determine the beam forming characteristics of the antenna array in our UAV communication system.












\setlength{\textfloatsep}{0pt}
\begin{figure*}[t!]
\centering
\includegraphics[scale=.8]{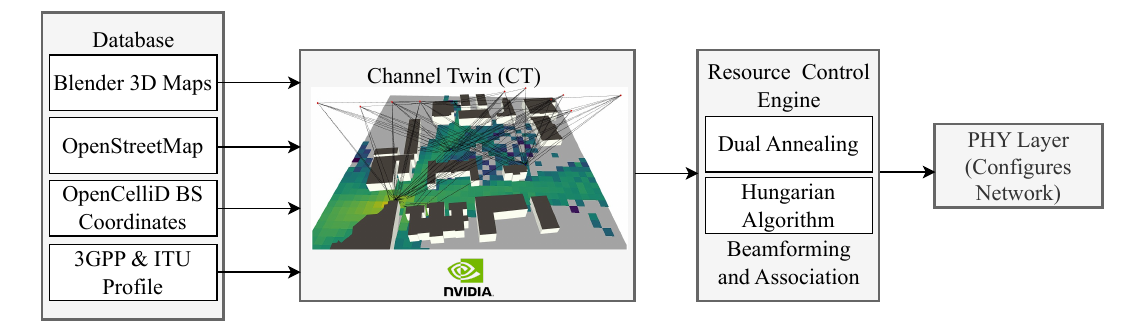}
\caption{Proposed Drone-BS-beam Association System Architecture}
\label{fig2}
\end{figure*}
\setlength{\textfloatsep}{0pt}

\section{Proposed Drone-BS Association and Resource Allocation Algorithm}

\subsection{Problem Formulation}
We first calculate the signal-to-interference plus noise ratio (SINR) of the $m$-th drone assuming it is scheduled to the $r$-th radio resource block (RRB) of the $l$-th BS. 
Assume that the drone's angular position is $(\theta_m, \phi_m)$. To formulate the optimization problem, 
%
we introduce two binary indicator variables, $\beta_{m,l}$ and $x_{m,l}^{(n)}$, where $\beta_{m,l} = 1$ if the $m$-th drone is associated with the $l$-th BS, and $0$ otherwise; and $x_{m,l}^{(n)} = 1$ if the $m$-th drone is assigned to the $n$-th beam of the $l$-th BS, and $0$ otherwise.
%
The received downlink SINR at the $m$-th drone from the associated BS and scheduled RRB  is expressed as
\begin{equation}\label{SINR_equation}
    \text{SINR}_{m,l}^{(r)}= \frac{ \sum_{n=1}^N \beta_{m,l} x_{m,l}^{(n)} P |h_{m,l}|^2 10^{\frac{G_{5G}(w_{n,l}\theta_{m,l},\phi_{m,l})}{10}}}{\sum_{l'=1, l' \neq l}^L I_{l'}^{(m)} +\sigma^2}.
\end{equation}
Here, $P$ denotes the transmit power of each base station, $|h_{m,l}|^2$ is the CT computed time-varying channel gain between the $m$-th UAV and the $l$-th BS, and $G_{5\mathrm{G}}(w_{n,l}, \theta_{m,l}, \phi_{m,l})$ represents the directional antenna gain in dB for beam $n$ of BS $l$ toward UAV $m$, computed using the 3GPP antenna model. Moreover, $\sigma^2$ represents the noise power and $I_{l'}^{(m)}$ denotes the interference power received at the $m$-th UAV from an interfering BS $l' \neq l$ defined as 
\begin{equation}
    \label{Interfernece}
    I_{l'}^{(m)}=\sum_{ \substack{m'=1 \\m' \neq m}}^M \sum_{n=1}^N \beta_{m,l'} \alpha_{m,l'}^{(r)}  x_{m,l'}^{(n)} P |h_{m,l'}|^2 10^{\frac{G(w_{n,l'},\theta_m,\phi_m)}{10}}.
\end{equation}
The received throughput at the $m$-th drone can be expressed as
\begin{equation}
    \label{rate}
    \mathrm{R}_m=W \sum_{l=1}^L \sum_{r=1}^R \log_2\left( 1+ \text{SINR}_{m,l}^{(r)}\right).
\end{equation}
Considering all the parameters, the optimization problem is formulated by
\begin{equation}
			\label{Opt_P0}
			\begin{split}
				\text{(P0)} & \max_{\bm{\beta, x} \in \{0,1\}} \mathrm{R}_m\\
				&\hspace{-0.8cm} \text{s.t.} \begin{cases}
				& \text{C1:}  \sum_{l=1}^L \beta_{m,l} =1, \forall m \\
                & \text{C2:}  \sum_{m=1}^M \beta_{m,l} \leq N, \forall l\\
                & \text{C3:}  \sum_{l=1}^L\sum_{n=1}^N \beta_{m,l} x_{m,l}^{(n)} =1, \forall m\\
                  & \text{C4:}  \sum_{l=1}^L\sum_{m=1}^M \beta_{m,l} x_{m,l}^{(n)} \leq 1, \forall n\\  
                    	\end{cases}
			\end{split}
		\end{equation}
Constraint C1 stipulates that each UAV can be associated with at most one base station (BS). Constraint C2 limits the load per BS, allowing any given BS to serve no more than 
$N$ UAVs simultaneously. Constraint C3 enforces beam-level exclusivity within a BS, requiring every UAV to operate on exactly one beam. Finally, Constraint C4 ensures the reciprocal one-to-one mapping by allowing each beam at a BS to be assigned to at most one UAV.

$\text{P0}$ is provably NP-hard, as it involves maximizing a sum of ratio functions over integer variables. Moreover, solving this problem requires global CSI acquisition for all drones, beams, and BSs, which significantly increases signaling overhead. To address these challenges, we propose a DT-enabled two-stage framework: the first stage tunes the scan angle for each UAV–BS–beam configuration to maximize directional antenna gain, while the second stage leverages these optimized gains to determine the UAV–BS–beam associations that yield the highest overall system performance.
\subsection{Proposed Two-Stage Optimization Framework}
Fig.~\ref{fig2} illustrates our proposed DT-enabled optimization framework. To ensure optimal link quality, each UAV–BS–beam configuration must be carefully examined, as beam direction significantly influences array gain due to the highly directional nature of 5G antenna patterns. Therefore, selecting the most suitable scan angle for every UAV–BS–beam triplet is essential for maximizing signal strength and improving overall throughput. However, this problem is mathematically intractable and non-convex due to the complex beam pattern equations and multi-dimensional search space. To this end, we adopt a meta-heuristic approach using dual annealing, which combines the global exploration of simulated annealing with the local refinement of deterministic search, followed by UAV-BS-beam association, enabling convergence to robust UAV-BS-beam connectivity solution.\\
\noindent\textbf{Stage 1: Beam Gain Optimization via Dual Annealing}: In the first stage, the objective is to determine, for each beam in the link between a 5G BS and a UAV, the scan angle $\phi_{\text{scan}}$ that maximizes the gain defined in \eqref{antenna_gain} by solving the optimization problem: 
\begin{equation}
\phi_{\text{scan}}^{\star} = \arg\max_{\phi_{\text{scan}} \in [-\pi,\pi]
} \left( G_{5\mathrm{G}}(\theta, \phi_{\text{scan}} ) \right).
\end{equation}
This non-convex optimization is solved using the \textit{dual annealing} algorithm, which performs global optimization by combining simulated annealing with local search. For each UAV–BS–beam combination, the scan angle $\phi_{\text{scan}}$ is optimized within the domain $[-\pi, \pi]$, and the resulting beam gain is stored for the subsequent assignment stage.The algorithm operates as follows:
\begin{enumerate}
    \item \textbf{Initialization:} Randomly initialize $\phi_{\text{scan}}$ within a bounded domain ($[-\pi, \pi]$).
    \item \textbf{Simulated Annealing Phase:} Explore the objective function landscape using probabilistic transitions that allow uphill moves, helping to escape local minima.
    \item \textbf{Local Search Phase:} Once a promising region is identified, a deterministic local optimizer refines the solution to find a local maximum.
    \item \textbf{Repeat:} The global and local search phases are alternated until convergence or a maximum iteration limit is reached.
\end{enumerate}
By combining global exploration with local exploitation, dual annealing ensures a high probability of finding the globally optimal beam direction for each UAV–BS–beam link.

\noindent\textbf{Stage 2: UAV–BS–Beam Association via Hungarian Algorithm:} In the second stage, the goal is to assign each UAV to a unique BS and one of its available beams, such that the total received signal power across all UAVs is maximized. We define the effective received power at the $m$-th UAV from the $l$-th BS utilizing $n$-th beam from \eqref{SINR_equation} as follows:
\begin{equation}
\Lambda_{m,l,n} = P  |h_{m,l}|^2  10^{\frac{G_{5G}(w_{n,l}, \theta_m, \phi_m^*)}{10}}.
\end{equation}
The joint optimization problem is then formulated as follows:
\begin{equation}
\label{Opt_P1}
\begin{split}
\text{(P1)} \quad & \max_{\bm{\beta}, \bm{x} \in \{0,1\}} \sum_{m=1}^M \sum_{l=1}^L \sum_{n=1}^N \beta_{m,l}  x_{m,l}^{(n)}  \Lambda_{m,l,n} \\
\text{s.t.} \quad & \text{Constraints C1 -- C4}.
\end{split}
\end{equation}
To solve this discrete linear sum assignment problem (LSAP) efficiently, we employ the \textit{Hungarian algorithm}, a polynomial-time method to solve linear sum assignment problems. The LSAP seeks to assign a set of tasks, in this case, UAVs, to a set of BS-beam pairs so that the total cost is minimized or, equivalently, the total utility is maximized. In this framework, the utility is defined by the effective link gain matrix $\Lambda \in \mathbb{R}^{M \times L \times N}$. To apply the \textit{Hungarian algorithm}, $\Lambda$ is reshaped into a two-dimensional cost matrix of size $M \times (LN)$, and its values are negated to convert the maximization problem into a minimization one. The \textit{Hungarian algorithm} proceeds through the following steps:
\begin{enumerate}
    \item \textbf{Cost Matrix Construction:} Flatten the 3D gain matrix into a 2D cost matrix of shape $M \times (LN)$.
    \item \textbf{Row and Column Reduction:} Normalize the matrix by subtracting row-wise and column-wise minimum values to facilitate zero-based matching.
    \item \textbf{Zero Covering:} Identify a minimum number of rows and columns needed to cover all zeros in the matrix.
    \item \textbf{Optimal Matching:} Using uncovered zeros, the algorithm finds an optimal one-to-one assignment that minimizes the total cost.
\end{enumerate}
The resulting assignment is decoded into two binary matrices. $\beta[m,l] = 1$ if UAV $m$ is assigned to BS $l$ and $x[m,l,n] = 1$ if UAV $m$ uses the $n$-th beam of BS $l$. This stage ensures that each UAV is served by a unique, non-conflicting BS–beam pair, leading to efficient spatial resource utilization and enhanced system throughput. Afterwards, the resulting optimized $\beta[m,l]$ and $x[m,l,n]$ is utilized by Eq. \eqref{SINR_equation} and Eq. \eqref{rate} respectively to calculate the optimized throughput achieved at UAVs.

\subsection{Computational Complexity}
\noindent
The dual--annealing search in \textbf{Stage~1} evaluates every UAV--BS--beam triplet, yielding a computational cost of $\mathcal{O}\!\bigl((T_{\mathrm{g}}+T_{\ell})\,M\,L\,N\bigr)$, where $T_{\mathrm{g}}$ and $T_{\ell}$ are the global and local iteration counts, respectively. \textbf{Stage~2} reshapes the three--dimensional link--gain tensor into an $M\times(LN)$ cost matrix and solves the resulting linear--sum assignment with the Hungarian algorithm, which runs in $\mathcal{O}\!\bigl(\max\{M,LN\}^{3}\bigr)$ time. 
Accordingly, runtime is dominated by Stage~1 when $(T_{\mathrm{g}}+T_{\ell})N \gg \max\{M,LN\}^{2}$; otherwise, the cubic complexity of the Hungarian step becomes the primary bottleneck. 
Both stages share the same memory footprint of $\mathcal{O}(M L N)$ complex entries required to store the channel tensor and the beam--gain table, ensuring that the framework remains tractable.


\begin{table}[t]
\centering
\vspace{0.3cm}
\caption{Simulation Parameters}
\label{tab:sim_params}
\begin{tabular}{|l|c|}
\hline
\textbf{Parameter} & \textbf{Value} \\
\hline
Carrier Frequency & 3.5 GHz \\
\hline
Bandwidth ($W$) & 30 MHz \\
\hline
Transmit Power ($P$) & 10 W \\
\hline
Noise Power ($\sigma^2$) & 0.3 W \\
\hline
Number of BSs & 4\\
\hline
Number of BS Antennas & $4\times 4$ \\
\hline
Number of UAV Antennas & $1\times 1$ \\
\hline
Beam Codebook Size ($N$) & 16 \\
\hline
Antenna Spacing ($d_h$, $d_v$) & 0.5$\lambda$ \\
\hline
Number of Drones ($M$) & 10, 20, 30, 40 \\
\hline
UAV Altitude & 100 m \\
\hline
UAV Distribution & Circular \\
\hline
Scene Resolution (HF) & $10^6$ rays \\
\hline
Scene Resolution (LF) & 100 rays \\
\hline
Rician $K$ Factor & 3 dB \\
\hline
Antenna Pattern & 3GPP Sectorized \\
\hline
Vertical Tilt Angle & 15$^\circ$ \\
\hline
Max Element Gain ($G_{E,\text{max}}$) & --8 dBi \\
\hline
Elevation Beamwidth ($\theta_{3\text{dB}}$) & 65$^\circ$ \\
\hline
Azimuth Beamwidth ($\phi_{3\text{dB}}$) & 90$^\circ$ \\
\hline
Front-to-Back Ratio ($A_m$) & 30 dB \\
\hline
Side-lobe Level Limit ($SL_{A,V}$) & 30 dB \\
\hline
\end{tabular}
\end{table}

\section{Simulation Results}
In this section, we evaluate the performance of the proposed CT-assisted resource allocation scheme and show their comparative effectiveness with pertinent baseline schemes. A summary of key simulation parameters is provided in Table~\ref{tab:sim_params}.
\noindent\textbf{Proposed Scheme:} In this proposed approach, we generate channel gains with a high-fidelity CT (HF-CT) setup. The HF-DT employs $10^6$ rays to model the channel impulse response (CIR), capturing fine-grained multipath propagation effects. We assume that this high-resolution ray tracing closely approximates real-world channel conditions. 
\setlength{\textfloatsep}{0pt}
\begin{figure}[h]
\centering
\includegraphics[scale=.28]{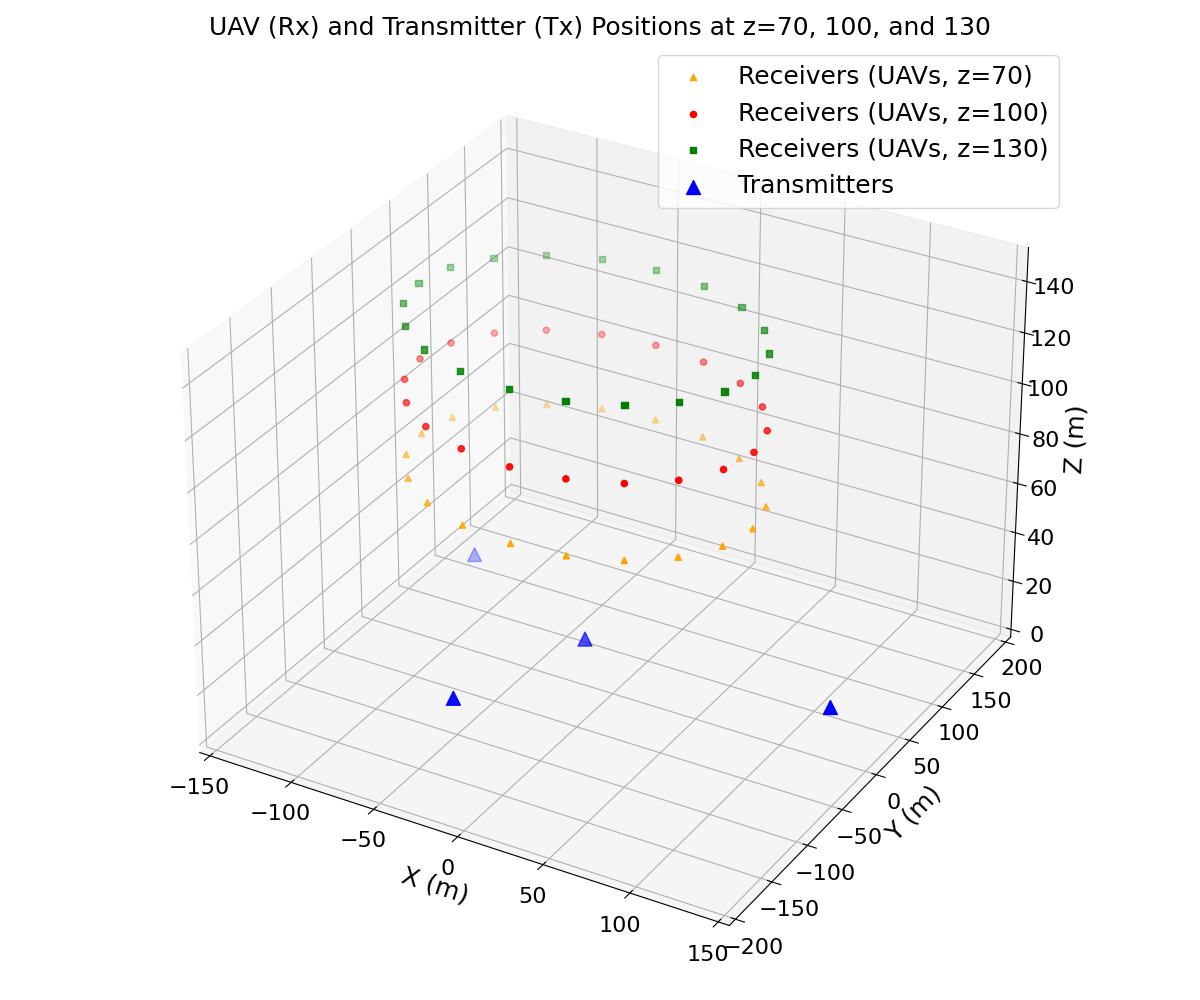}
\caption{RX TX location visualization}
\label{fig:uav_tx_positions}
\end{figure}


\begin{table}[ht]
\centering
\caption{Average Throughput (Mbps) at Varying UAV Altitudes}
\label{tab:avg_tput_vs_altitude}
\begin{tabular}{l|c|c|c}
\hline
\textbf{Method} & \textbf{Sc 1 (Mbps)} & \textbf{Sc 2 (Mbps)} & \textbf{Sc 3  (Mbps)} \\
\hline
Sionna (RT) HF & 324.66 & 187.35 & 107.90 \\
\hline
Sionna (RT) LF & 282.93 & 166.93 & 100.95 \\
\hline
Model-based CT & 289.92 & 161.73 & 100.57 \\
\hline
Random         & 175.45 & 61.42  & 31.54  \\
\hline
Closest BS     & 126.57 & 36.13  & 13.24  \\
\hline
\end{tabular}
\end{table}
\setlength{\textfloatsep}{0pt}
\begin{figure}[h]
\centering
\includegraphics[scale=.57]{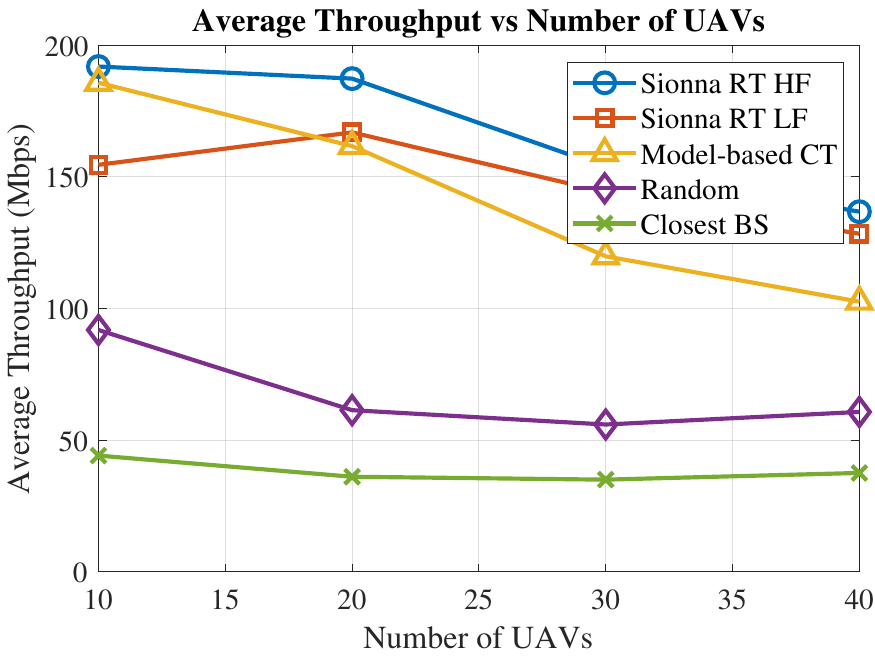}
\caption{Throughput Comparison for various UAVs}
\label{result2}
\end{figure}
\setlength{\textfloatsep}{0pt}
\begin{figure}[h]
\centering
\includegraphics[scale=.57]{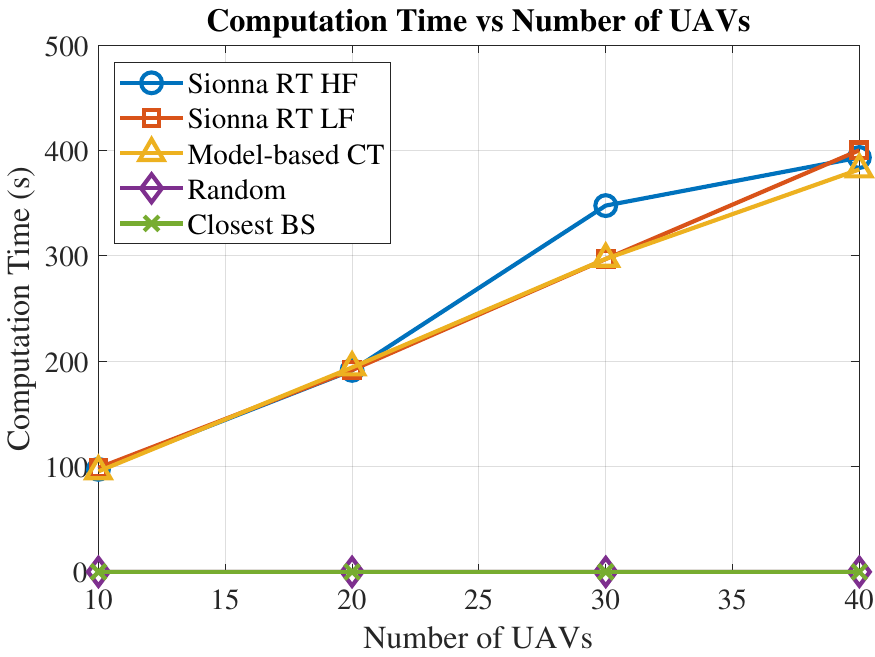}
\caption{Computation Time Comparison for various UAVs}
\label{result3}
\end{figure}

\subsection{Baseline Schemes for Comparison}
\noindent\textbf{Baseline Schemes for Comparison:} To benchmark our method, we compare it against four baseline schemes. Importantly, for fairness, throughput in all baselines is evaluated using HF-CT CIRs as they present a close-to-real-world channel. This allows us to assess how well each scheme would perform under real-world channel conditions. It is worth mentioning that Baseline Schemes 1 and 2 assist the proposed scheme in demonstrating the effectiveness of using a highly precise CT. Likewise, Baseline Schemes 3 and 4 emphasize optimizing the resources to achieve higher system throughput.

\noindent\textit{\textbf{Baseline 1: Low-Fidelity CT (LF-CT)}}: In this baseline, channel gains are generated using a LF CT that models the CIR using only 100 rays in Sionna. This offers a lightweight alternative with significantly reduced complexity, but at the cost of lower accuracy to estimate CIRs. Resource allocation is carried out using the Hungarian method based on these LF-DT gains. Throughput is then computed using the HF-DT channel, allowing us to examine how reduced model fidelity affects real-world performance.

\noindent\textit{\textbf{Baseline 2: Model-based CT}}: Here, we use a 3GPP Rician fading model with Urban Microcell (UMi) path loss to generate channel gains. This statistical model does not rely on environment-specific ray tracing and serves as a standard benchmark in wireless simulations. Resources are optimized using the Hungarian method based on these statistical gains, and throughput is evaluated using the HF-DT.

\noindent\textit{\textbf{Baseline 3: Random Assignment}}: In this naive strategy, UAVs are assigned randomly to base stations and beams, with the constraint that no beam is reused. No channel gain or beamforming information is considered. Throughput, again, is evaluated using the HF-DT channel.

\noindent\textit{\textbf{Baseline 4: Closest BS with Best Beam}}: In this heuristic, each UAV is first assigned to its nearest base station based on Euclidean distance. From the available beams at that BS, the one with the highest effective gain ($\Lambda_{m,l,n}$) is selected. This method incorporates limited spatial awareness without performing global optimization. Similarly, based on the optimization assignment, throughput is evaluated using the HF-DT.

\subsection{Performance Analysis}
For performance evaluation, we define the spatial distribution of UAVs and BSs within a DT of Howard University campus constructed in Blended from real 3D building models obtained via the OpenCellID database. 

Fig.~\ref{fig:uav_tx_positions} illustrates the 3D spatial arrangement of 20 UAVs and BSs in three representative drone-corridor configurations. In Scenario 1 (Sc 1), the UAVs (yellow dots) traverse a circular flight path at an altitude of 75 m. Scenario 2 (Sc 2) positions the UAVs (red dots) on a circular trajectory at 100 m, while Scenario 3 (Sc 3) places them (green dots) on an analogous path at 130 m. The BS transmitters with their locations taken from OpenCellID database are shown as blue triangles. 

The corresponding average throughputs are summarized in Table \ref{tab:avg_tput_vs_altitude}. Across all scenarios, the proposed HF-CT-aided resource-allocation scheme consistently surpasses the baseline methods in achieving better throughput. For example, at 100 m altitude, the HF-CT scheme yields throughput gains of 10.9\%, 13.7\%, 67.2\%, and 80.7\% relative to the LF-CT, 3GPP, random, and closest-BS schemes, respectively—highlighting the utility of an accurately designed CT and optimized resource allocation in drone corridors. As the BS–UAV separation grows, the resulting path-loss increase degrades overall performance, and the relative advantage of the proposed approach narrows. Specifically, raising the flight altitude from 75 m to 100 m reduces the performance gap between the HF-CT and LF-CT schemes from 41.73 Mbps to 20.42 Mbps; elevating the altitude further to 130 m decreases this gap to 6.95 Mbps.

Fig.~\ref{result2} illustrates how the network responds to progressive densification of drones in the corridor by comparing the proposed approach with all baseline schemes. A monotonic decline in average network throughput is evident across all methods, attributable to rising levels of interference as the number of active UAVs increases. Throughout this range, the HF-CT scheme consistently outperforms every baseline, whereas the LF-CT and the statistical 3GPP Rician scheme exhibit poor performance. These findings emphasize that accurate environmental knowledge provided by HF-CT not only elevates absolute throughput but also preserves a substantial performance margin over competing schemes as user density grows.

Fig.~\ref{result3} quantifies the computation-time overhead associated with the four resource-allocation strategies. For the HF-CT- and LF-CT-based schemes, processing time rises with the number of UAVs, reaching about 393 s when 40 UAVs are present; this growth reflects the increasing complexity of the underlying optimization. By contrast, the random and closest-BS heuristics complete almost instantaneously because they bypass any optimization routine. Their speed, however, comes at the expense of markedly lower UAV throughput, underscoring the classic trade-off between computational efficiency and performance.




\vspace{-0.3cm}
\section{Conclusions}
In this paper, we proposed a novel high-fidelity CT-driven drone-BS-beam association algorithm for a drone corridor environment. The proposed algorithms comprise a two-stage optimization, which was solved using Dual Annealing and the Hungarian algorithm. The first algorithm optimizes the maximum antenna gain, while the latter formulates the optimal association between drone, BS, and beams. We compared the performance of our proposed scheme with baselines LF-CT, model-based CT, random assignment, and closest BS with best beam and evaluated the performance in terms of average throughput achieved at the UAVs. The simulation results concluded that the proposed optimization framework outperforms in terms of data rate for different system parameters.

\vspace{-0.1cm}
\section*{Acknowledgments}
This work was supported in part by the NSF Grant \# 2200640, in part by DoD/US Army Contract W911NF-22-1-022, and in part by the US DoD Center of Excellence in AI/ML at Howard University under Contract W911NF-20-2-0277 with the US ARL. However, any opinion, finding, and conclusions or recommendations expressed in this document are those of the authors and should not be interpreted as representing the official policies, either expressed or implied, of the funding agencies.

\balance
\bibliographystyle{IEEEtran}
\bibliography{references_TeamB}
\balance

\end{document}